\documentclass{sig-alternate}

\begin{document}
%
\conferenceinfo{ACM RS}{2012 Dublin, Ireland}

\title{Social Recommender Systems Based on Coupling Network Structure Analysis
}
%
%
%
%
%

\numberofauthors{4} 
%
\author{
%
%
\alignauthor
Xiao Hu\\
       \affaddr{Web Sciences Center, University of Electronic Science and Technology of China}\\
       \affaddr{Chengdu, P. R. China}\\
       \email{huxiao64@gmail.com}
\alignauthor
ChuiBo Chen\\
       \affaddr{School of Mathematical Sciences, University of Electronic Science and Technology of China}\\
       \affaddr{Chengdu, P. R. China}\\
       \email{chenchuibo@gmail.com}
\and
\alignauthor
Xiaolong Chen\\
       \affaddr{School of Mathematical Sciences, University of Electronic Science and Technology of China}\\
       \affaddr{Chengdu, P. R. China}\\
       \email{chenxiaolong115@gmail.com}
\alignauthor
Zi-Ke Zhang\\
       \affaddr{Institute of Information Economy, Hangzhou Normal University }\\
       \affaddr{Hangzhou, P. R. China}\\
       \email{zhangzike@gmail.com}
}
\date{9 April 2012}

\maketitle
\begin{abstract}
The past few years has witnessed the great success of recommender systems, which can significantly help users find relevant and interesting items for them in the information era. However, a vast class of researches in this area mainly focus on predicting missing links in bipartite user-item
networks (represented as behavioral networks). Comparatively, the social impact, especially the network structure based properties, is relatively lack of study. In this paper, we firstly obtain five corresponding network-based features, including user activity, average neighbors' degree, clustering coefficient, assortative coefficient and discrimination, from social and behavioral networks, respectively. A hybrid algorithm is
proposed to integrate those features from two respective networks. Subsequently, we employ a machine learning process to use those features to provide recommendation results in a binary classifier method. Experimental results on a real dataset, \emph{Flixster}, suggest that the proposed method can significantly enhance the algorithmic accuracy. In addition, as network-based properties consider not only the social activities,
but also take into account user preferences in the behavioral networks, therefore, it performs much better than that from either social or behavioral networks. Furthermore, since the features based on the behavioral network contain more diverse and meaningfully structural information, they play a vital role in uncovering users' potential preference, which, might show light in deeply understanding the structure and function of the social and behavioral networks.
\end{abstract}




\maketitle
\section{Introduction}
Network sciences have provided us powerful and versatile tools to well recognize and understand various systems, ranging from economic systems to human society, from computer sciences to biology, and so on \cite{barabasi1999emergence,newman2010networks}. There is a vast class of works in this area mainly focusing on independently studying the \emph{Social Networks} and the \emph{Behavioral Networks}. Social Networks, composed of identical individuals, describe the properties of essential components and various kinds of interactions among them~\cite{jamali2006different}. Behavioral networks, on the other hand, consisted of two different components, users and items, involve not only the user profiles and item properties, but also contain user-item behavioral records. Consequently, they constitute a so-called \emph{Bipartite Graphs}, on which researchers have devoted much effort to predict user potential preferences via mining the binary relations~\cite{XiangL201001}. However, although social networks have been realized to play important roles in uncovering user hidden interests, the majority of works on this topic mainly concentrated on making use of the trustworthiness among individuals to make better recommendation results~\cite{ziegler2005,nie2011}. During the last decade, with the rapid growth and wide application of \emph{Social Network Services} (SNS) (e.g. \emph{Twitter.com}, \emph{Facebook.com}, \emph{Weibo.com}, etc.), users in those platforms can not only keep social contacts with friends, but also can share common interests, such as movies, stories, music and so on. That is to say, such social activities can benefit individuals in effectively finding more interesting items. Furthermore, those common interests/behaviors can conversely enhance the social relationships and help users building new relationships with congenial friends. Therefore, there is huge overlap between social and behavioral networks, in which the users play as coupling nodes, which drive the interdependent  \cite{gao2011networks} or interconnected \cite{yeh2002virtual} properties and related functions.

Recently, \emph{Recommender Systems}, aiming at helping users find relevant and interesting items from the information era for them. The most widely adopted framework is the classical \emph{Collaborative Filtering} (CF) \cite{resnick1994grouplens,sarwar2001item}, which provides the recommendation by taking into account effects of the most similar users/items. Nowadays, many pioneering works have tried to apply social factors into the CF framework~\cite{kautz1997referral, konstas2009social, nie2011}. However, most of them focused on estimating the similarities among users, respectively from social and behaviors, and then hybriding the two metrics in order to obtain better recommendation performance. Despite its effectiveness, the network structure, which contains rich information of users' social and behavioral preferences, is relatively lack of attention. Actually, such network structure has common features for both social and behavioral networks, e.g. active users are more likely to make acquittance with more friends, as well as collecting more items. And similar users attend to like similar movies. All these aspects can be described and analyzed by network structure properties. In this paper, we discuss this issue via building a machine learning model. Basically, the research problem that we are interested in, is related to following points:

\begin{itemize}
  \item { }
  How we simultaneously define the corresponding structural features from social and behavioral networks?
  \item { }
  With a binary classifier, how we evaluate the correlations and differences between social and behavioral networks?
\end{itemize}

We firstly obtain five corresponding network-based features, including user activity, average neighbors' degree, clustering coefficient, assortative coefficient and discrimination, from social and behavioral networks, respectively. We then employ a machine learning process to use those features to provide recommendation results in a binary classifier method with ensemble learning. Experimental results on a real dataset, \emph{Flixster}, suggest that it can significantly enhance the algorithmic accuracy.

This paper is organized as follows. Section 2 is about the related works on coupling networks and recommender systems. In Section 3, we describe the extracted features and the proposed model. Section 4 describes the data and shows experimental results. In Section 5, we conclude our work and discuss the future works.

\section{Related Work}
\subsection{The Coupling Networks}
A large number of measures have been proposed to characterize network properties. Examples include the clustering \cite{watts1998collective}, network modularity \cite{newman2006modularity}, and degree correlation~\cite{fowler2009model}, etc. These measures demonstrate that real networks indeed involve meaningful local structures (e.g. Network motifs \cite{chen2006nemofinder}), and provide useful information for various applications.

The coupling networks, also known as interdependent networks, normally contain a two-layer network, such as electricity and Internet networks \cite{buldyrev2010catastrophic}, airport and railway networks \cite{givoni2006airline}. There is a kind of coupling nodes, such as cities in the two aforementioned networks, which play the interconnection and maintenance roles between these two-layer networks. Consequently, those nodes are critically important for the robustness of whole networks \cite{gao2011networks}. Social networks, similar with the interdependent networks, also involve such coupling nodes, users, who both make friends and collect items. Therefore, those users are especially vital for maintaining the structure, connectivity and robustness of social and behavioral networks.

\subsection{Social Recommender Systems}
The past few years have witnessed the great success of recommender systems, which can significantly help users find relevant and interesting items for them in the information era \cite{resnick1997recommender}. Recently, many efforts have been devoted to study social networks to explore collaborative interactions or influences in order to produce more reliable recommendations \cite{huang2012exploring}. \cite{jamali2006different} suggested a mixture between social influence and similarity of activities to predict future behaviors. \cite{esslimani2009social} presented a measurement analysis of various online social networks by studying the corresponding topological properties, and then make recommendations. Collaborative filtering based on social networks, wherein user-selected like-minded alters are were to make predictions, outperformed traditional user-based CF in predictive accuracy \cite{zheng2008social}. \cite{esslimani2009social} proposed a new behavioral network based CF, exploited navigational patterns and transitive links to model users, analyzed behavior similarities, and eventually explored missing links.

In addition, \emph{Machine Learning} has been proved to be a very useful tool in enhancing the recommendation accuracy \cite{chen2008combinational}. \cite{breese1998empirical} described several algorithms to use user preferences to predict additional topics, including techniques based on correlation coefficients, vector-based similarity calculations, and statistical Bayesian methods. \cite{paparrizos2011machine} solved the problem of seeking new jobs with a supervised machine learning method. \cite{knees2011towards} proposed to combine rule patterns and supervised learning to extract semantic music information.

However and to the best of our knowledge, there is no previous work on using machining learning process to make recommendations from coupling network structure features, which is the main contribution of this paper and will be presented next in detail.

\section{Features and Model}

For the observed social network, we firstly extract the largest connected sub-graph, in which every node represents a user, and each edge is undirected. Then, we refine the behavioral network to guarantee that the two networks have the same group of users. Finally, we make an interdependent network consisted of this social and behavioral network, and then extract the features and transfer it to a binary classification problem. In particular, we choose five representative features, including user activity, average degree, clustering coefficient, assortative coefficient and discrimination, defined as following:

\begin{itemize}
  \item {User Activity ($A$)} -- it refers to the total number of user $u_i$'s friends in social network, while in behavioral network it means the number of movies watched by $u_i$.
  \item {Average Activity of Neighbors ($AN$)} --  defined as $AN_i = \frac{\sum\limits_{j\in \Gamma_{i}}A_j}{A_i}$, where $\Gamma_{i}$ is the set of user $u_i$'s neighbors. In social network, it is $u_i$'s friends list, and it refers to all $u_i$'s watched movies in the behavioral network.
  \item {Clustering Coefficient ($C$)} -- For social networks, it is defined as $C_i=\frac{\sum_{j,m}a_{ij}a_{jm}a_{mi}}{A_i(A_i-1)}, \label{E03}$ where $a_{ij}$ = 1 if node $i$ is connected to node $j$, otherwise $a_{ij}$ =0. Thus, $C_i$, in social networks, expresses the likelihood for two neighbors $j$ and $m$ of node $i$ are connected; for behavioral networks, it is defined as $C_i= \frac{\sum\limits_{j \neq j}{s_{jk}}}{A_i*(A_i-1)}$, where $s_{jk}$ is the Jaccard index~\cite{jaccard1901}. $C_i$, in behavioral networks, expresses how similar the user $u_i$'s movies are. It also suggests how diverse $u_i$'s taste is.
  \item {Assortative Coefficient ($AC$) } -- defined as \\ $AC_i= \frac{A_i^{-1}\sum\limits_{j\in\Gamma_{i}}{A_{i}A_j-[A_i^{-1}\sum\limits_{j\in\Gamma_{i}}\frac{1}{2}(A_i+A_{j})]^2}}{          A_i^{-1}\sum\limits_{j\in\Gamma_{i}}\frac{1}{2}(A_i^2+A_{j}^2)-[A_i^{-1}\sum\limits_{j\in\Gamma_{i}}\frac{1}{2}(A_i+A_{j})]^2}$, and $AC_i \in [-1,1]$. $u_i$ is more likely to connect reversely active users (or popular items in behavioral networks) if $AC_i<0$, and vice verse.
  \item {Discrimination ($D$)} -- defined as $D=1-\sum\limits_{A_j}{p^2(A_j)}$, where $A_j$ runs over all possible activities of $u_i$'s neighbors and $p(A_j)$ is the ratio of $A_j$-activities to total different activities of user $u_i$. If $u_i$ connects to users (or items) with only one kind of activity (e.g. $A_j=10$), $D_i=0$. The larger $D$, the more discriminating of user $u_i$'s neighboring activities.
\end{itemize}

We then employ the machine learning process to evaluate the correlations and differences between social and behavioral networks. Given a set of users, each is described by the defined features (each feature is normalized) as the input variables, and labeled to indicate whether this user and its friends in the social network have watched randomly selected movies. We then adopt the \emph{c4.5} to be the weak classifier and the AdaBoost \cite{freund1995desicion} to form the strong classifier. To better investigate effects of social and behavioral features, we introduce a tunable parameter, $\alpha \in [0,1]$, where $\alpha$ refers the social impact, and $1-\alpha$ simultaneously corresponds to the impacts of behavioral features. In the extremal cases $\alpha=0$ and $\alpha=1$, features are taken into account only from social and behavioral networks, respectively. Subsequently, we apply the decision tree algorithm to weighted training data and then take the classifier with the highest score as the strong classifier, and eventually use it to predict testing instances.

\section{Data and Results}
In this paper, one representative dataset, \emph{Flixster.com}, a social movie rating website (with five-star rating level), is used to evaluate the proposed method. We purify the data in order to remove noisy data by: (i) extracting the largest connected sub-graph from the social network; (ii) deleting the absent users from the behavioral network to guarantee all users to be existing in the social network, and we delete the movies which are never watched by the users in the social network. Finally, we get the social network with 137,372 users and 2,430,282 undirected links, and the behavioral network with the same number of users, 48,756 movies, and 8,062,853 user-item-rating behavioral records. For the binary classifier, we divide users into two classifiers: we mark 1 to the users who and friends have watched randomly selected popular movies with rating larger than 3, and 0 otherwise.

\begin{figure}
\centering
  \includegraphics[width=9cm]{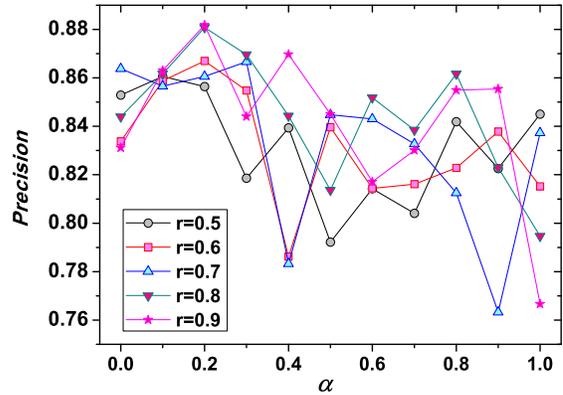}
  \caption{Precision (recommendation accuracy) versus $\alpha$. Each line represents the result of one dividing ratio of training and testing data (denoted as $r$).} \label{results}
\end{figure}

Figure \ref{results} shows the recommendation accuracy (we use \emph{Precision} in this paper) for $\alpha \in [0,1]$. It is can be seen that, for different ratios of training data, the optimal result reaches at $\alpha = 0.2$. Considering that $\alpha$ indicates the impact of features of social network structure, it suggests that, for \emph{Flixter}, the behavioral network is more significant than the social network in predicting users' potential-like movies. The optimal $\alpha>0$ also indicates that the social features indeed can improve accuracy of recommendations.

\section{conclusions and Discussion}
In this paper, we have proposed to use network structure based features of coupling networks, and combine a machine learning process to evaluate the performance of the extracted properties in recommendation accuracy. The experimental results show that the behavioral information plays more important role in the observed data, \emph{Flixter}. In addition, the results also shows that the social features can benefit in improving recommendations accuracy.

This article only provides a simple start point for the design of hybrid algorithms making use of social information, while a couple of issues remain open for future study. Firstly, we lack quantitative understanding of the structure and dynamics of coupling networks. Although the relation between the coupling property and recommender systems is not clear thus far, we deem that an in-depth understanding of coupling networks should be helpful for better recommendations; Secondly, the current algorithm only considers the effects of five representative network-based features, and more meaningful features should also be taken into account, thus the Social Network Analysis (SNA) based techniques can be used to provide more substantial recommendations, and social predictions as well. Thirdly, the experiment in this article is only performed on one dataset. For different coupling networks, the social impact should perform quite differently, according to their respective underlying structure. Finally, this paper provides a promising way to identify the importance of different features in driving the connectivity and robustness of various coupling networks.

\section*{Acknowledgments}

This work was partially supported by Natural Science Foundation of China (Grant Nos. 11105024, 61103109, 60973069).

\bibliographystyle{abbrv}
\bibliography{v4}
\end{document}